\documentclass[11pt,a4paper]{article}
\usepackage{amsmath, amssymb, verbatim, graphicx, color, bbm, framed, hyperref}
\usepackage{fullpage, changes}

\newcommand{\ket}[1]{| #1 \rangle}

\newcommand{\bra}[1]{\langle #1 |}

\newcommand{\proj}[1]{\ket{#1}\!\bra{#1}}

\begin{document}

\title{Certified randomness in quantum physics}
\author{Antonio Ac\'{\i}n$^{1,2}$ and Lluis Masanes$^{3}$\\[0.5em]
$^{1}$ICFO-Institut de Ciencies Fotoniques, The Barcelona Institute of Science\\ and Technology,
08860 Castelldefels (Barcelona), Spain\\
$^{2}$ICREA-Instituci\'o Catalana de Recerca i Estudis Avan\c
cats, 08010 Barcelona, Spain\\
$^3$Department of Physics \& Astronomy, University College of London,\\
London WC1E 6BT, United Kingdom}

\date{}

\maketitle

\begin{abstract}
%Randomness is a valuable resource in cryptography, randomized algorithms, and computer simulations.
%
%Standard methods for generating randomness rely on strong assumptions on the devices that are difficult to meet in practice. However quantum correlations allow for novel methods for generating random numbers relying only on minimal assumptions and, in particular, without requiring any assumption on the inner working of the devices. These methods are known as device-independent and open new avenues for the generation of randomness.

%In this review we provide an introduction to the field of device-independent random number generation, a description of its main achievements, and a list of its challenges. First, we explain the essential mechanism that allows to certify randomness from correlations that violate a Bell inequality, and how this can be achieved without any need for trusting the quantum devices used. Second, we describe the existing protocols and compare their performance and security. We also provide an overview of the experimental implementations and the difficulties  involved. And finally, we elucidate the implications that these experiments and theoretical results have for our understanding of nature, and the constraints that they impose on any future physical theory that may supersede the contemporary ones.
The concept of randomness plays an important role in many
disciplines. On one hand, the question of whether random processes
exist is fundamental for our understanding of nature. On the other
hand, randomness is a resource for cryptography, algorithms and
simulations.  Standard methods for generating randomness rely on
assumptions on the devices that are difficult to meet in practice.
However, quantum technologies allow for new methods for generating
certified randomness. These methods are known as
device-independent because do not rely on any modeling of the
devices. Here we review the efforts and challenges to design
device-independent randomness generators.
\end{abstract}

%\vspace{8mm} 

\section{Introduction}

%\subsection{Random numbers}

%Randomness is a fascinating concept that attracts the interest of different communities, from philosophers to mathematicians, physicists and computer scientists. On one hand, the question of whether random processes exist is fundamental for our understanding of nature. On the other hand, randomness is a valuable information resource in cryptography, randomized algorithms, and the simulation of physical systems.
%Randomness is a fascinating concept that attracts the interest of many different communities, from philosophers to mathematicians, physicists and computer scientists. On the one hand, the question of whether random processes exist is fundamental for our understanding of nature. On the other hand, randomness is a valuable information resource in many cryptographic scenarios, in reducing the complexity of computer algorithms, and the simulation of physical systems.

Because of its importance, a significant scientific effort is
devoted to understand when a given %physical
process generates ``good"
randomness. The process is represented by a device, or
black box, producing bits, see Figure~\ref{fig1}, which is in the
user's hands.
What constitutes ``good" randomness may depend on the
application, but here we are interested in the strongest
definition: $N$ bits are \emph{perfectly random} if they are unpredictable, not only to the user of the device, but to \emph{any observer}.
This definition is satisfactory both from a fundamental
and applied perspective. On the one hand, while unpredictability
by any observer may not be needed for some applications,
%requiring ``random" numbers,
such as Montecarlo simulations, from a
fundamental perspective it is difficult to argue that a process is
random if there could exist an observer able to predict its
outcomes. On the other hand, by demanding that the results should
look random to any observer, the generated randomness is
guaranteed to be \emph{private}: the user, by running the process
in a secure location, has the guarantee that nobody knows the
obtained results, which can later be safely used for cryptographic
purposes.

According to this definition, \emph{the generation of randomness
from scratch is impossible}. This follows from the unfalsifiable hypothesis of the existence of a super-deterministic
model in which everything, including all the history of our
universe, was pre-determined in advance and known by the external
observer. Thus, \emph{any protocol for randomness generation must be
based on some hypotheses or assumptions}. The appropriateness of the assumptions is often debatable and strongly depends on the
application. From a fundamental point of view, a random number
generator (RNG) is better than another if it is based on fewer or weaker
assumptions. However, adding more assumptions may not compromise
the use of the random bits for a specific application, while it
may simplify the experimental implementation and/or increase the
efficiency. %Sometimes, flipping a classical coin is enough,
%even if from a fundamental point of view this process is of
%deterministic nature \added{COM IMPOSES QUE LA MONEDA SIGUI NO QUANTICA?}\textbf{EL QUE VULL ES POSAR UN EXEMPLE QUE DEMOSTRA QUE EN CERTES SITUACIONS ACCEPTEM PROCESSES QUE PODEN PREDIR-SE. EN PODEM PARLAR, MAI M'HA AGRADAT MASSA AQUESTA FRASE}.

Here we adopt a physics-based approach to randomness generation: the random numbers
%the generated {symbols}
should be unpredictable to any
\emph{physical} observer,
%\footnote{One may wonder whether there exist non-physical observers, but we prefer to keep the distinction for the sake of clarity.\added{JO TREURIA AQUEST FOOTNOTE.}},
%given that he/she is
that is, any observer whose actions are constrained by the laws of physics.
In particular, and
%The two devices, of the user and \replaced{potential adversary}{ the external observer}, are physical and,
according to the current understanding of nature, the device generating the random numbers and the device held by the potential adversary should obey the laws of quantum physics.
%Here is a first assumption: any device is subjected to the laws of quantum physics.
Within this theory, the joint state of the user $U$
and the adversary $E$ (which includes all the environment) describing $N$ ideal random bits is
\begin{equation}\label{rndef}
    \rho_{UE}=
    \left(\frac{1}{2}\proj{0}+\frac{1}{2}\proj{1}\right)^{\!\otimes N}
    \hspace{-1mm} \otimes \rho_E\ .
\end{equation}
This corresponds to $N$ realizations of a perfect random %classical
bit, taking the value 0 and 1 with equal
probability, which are totally uncorrelated with the state of the environment $\rho_E$.
%According to this state, the external observer is unable to predict the user bits better than with a totally uninformed guess.
%Demanding that the output of a (quantum) physical RNG is state~\eqref{rndef} is an unnecessarily strong requirement.
Given a device, we say that it generates \emph{arbitrarily good randomness} if its output is undistinguishable from the ideal state~\eqref{rndef}, up to some controllable small error.

A fundamental issue when considering randomness generation is
certification: how can the user certify that the numbers produced
by his device are random? According to~\eqref{rndef}, the random bits should follow a uniform probability
distribution and also be uncorrelated to the
environment. Concerning the first point, the standard solution
consists of running statistical tests~\cite{nist} on
sequences generated by the device. However, it is unclear what
passing these tests means and, in fact, it is impossible to certify
with finite computational power that a given sequence is random.

The certification of privacy is much subtler. The best way to
illustrate this is by means of what we call the memory-stick
attack. Imagine a situation in which the provider of the devices is the adversary and has
access to a proper RNG. The provider uses it to generate a long
sequence of good random numbers, stores them into a memory stick
and sells it as a proper RNG to the user. The numbers generated by
the user will pass any statistical test
%\footnote{It is assumed that the memory is larger than the length of sequences used in the tests. \added{Jo treuria aquest footnote.}}
and look random. % from the user perspective.
However, they are not properly random, as they can be
perfectly predicted by the adversary. %While the memory-stick
%attack may seem a bit academical, it is a good exercise to
%confront any proposal for RNG against it.

%\subsection{Quantum random-number generators}

The generation of good randomness is a notoriously difficult
problem~\cite{vonneumann, sadhistory}. There exist basically three types of
approaches. Pseudo-random-number generators (PRNG) use an algorithm to process an initial random seed. They are fast,
cheap and the properties of the generated sequences are good
enough for some applications.
%However, it is clear that they cannot be considered as proper
%method for generation, as illustrated by the following famous
%quote by Von Neumann: ``Anyone who considers arithmetical methods
%of producing random digits is, of course, in a state of
%sin"~\cite{vonneumann}. They are also problematic for some
%applications~\cite{sadhistory}.
However, the random character of the output and their privacy is based on
%the assumption that the computational power of the adversary is limited.
assumptions on the computational power of the  adversary.
But this is not the criterion adopted here, as we demand
unpredictability to any observer, independently of its computational power.

The second type of RNG are called True RNG (TRNG) and exploit
physical processes that are hard to predict, such as
meteorological phenomena or the
mouse movements of a computer user. %A drawback of these solutions
%is that the random properties of the generated numbers crucially
%depend on the details of the implementation, which are often hard,
%or even impossible to control.
Finally, there are quantum RNG (QRNG), which exploit a quantum
process believed to be fundamentally random. %While it is preferable to make a
%distinction, QRNG can also be seen as an example of TRNG,
%using a quantum instead of a classical physical process.
In what follows, we focus this discussion on QRNG, although many of the problems stated below also apply to TRNG.

The paradigmatic example of a QRNG is defined by the clicks
observed after a single photon impinges a beam-splitter, see
Figure~\ref{fig1}. This is however an idealized theoretical
situation that may be difficult, if not impossible, to perfectly
meet in an experiment. Imperfections on the devices are
unavoidable and may deteriorate the quality of the generated
numbers in uncontrolled manners. It is also difficult to exclude
memory effects, for instance at the detectors, which produce correlations among the generated bits. The privacy of the
symbols follows from the fact that the single-photon state is
assumed to be pure and therefore cannot be coupled to another
system. Hence, there is plenty of assumptions on the working of the %components of the
QRNG that are crucial to guarantee the
perfect match between the ideal theoretical situation and the
implementation, which is in turn
necessary to guarantee the quality of the generated outputs.

The current certification method applied to QRNG consists of passing
statistical tests. Apart from the problems already mentioned above, the use
of these tests is even more doubtful for QRNG, as they can be
satisfied by classical RNG too. So, the only guarantee the user
has that the symbols have a quantum origin is trusting the
provider. %. In fact, as the user is never going to open the device
%and test the different components, he needs to trust not only that
%the provider has used quantum effects, but and the full quantum preparation.
Finally, the user has no means to test the privacy of
the symbols %. He gets a device from the provider and
as he is unable to rule out the memory-stick attack. Trust is again the
only solution. All this level of trust is unsatisfactory as
(i) in many situations, especially for cryptographic applications,
it is convenient to reduce the trust on the provider %\deleted{and the whole preparation process}
as much as possible, and (ii), even if the
provider is trusted and has constructed the devices in the best possible way, uncontrolled drifts and changes on the devices are unavoidable
and may deteriorate the quality of the generated randomness. There is
a need for solutions that certify the quantumness, %\added{TE SENTIT ``CERTIFICAR QUANTUMNESS"? TE ALGUNA UTILITAT?},
quality and privacy of QRNG without requiring any detailed modelling of the devices.

Device-independent quantum random-number generators (DIQRNG) offer a
solution to the previous issues and provide protocols for generating certified randomness
based only on general assumptions on the setup, such as, e.g., the validity of quantum physics. In particular,
they do not require any assumption on the inner working of the devices, which can be seen as quantum black boxes processing classical information. The development of DIQRNG protocols is an active
research field involving many concepts and methods, from information-theoretical studies to design stronger
protocols based on weaker assumptions, to their experimental realization using current or near-future technology.

In what follows, we first show how randomness certification without assumptions on the inner working of the devices can be achieved by exploiting the quantum violation of Bell inequalities (Section~\ref{diqrng}).
We then describe the state of the art in DIQRNG protocols (Section~\ref{protocols}) and their experimental implementations (Section~\ref{implementations}). %; and the challenges and future research directions of the field.
As these implementations turn out to be challenging, we describe other approaches to certified randomness with milder experimental requirements in Section~\ref{semidiqrng}. Finally, in Section~\ref{foundations} we explain how, in addition to practical applications, protocols for certified randomness answer some fundamental questions in physics. We conclude with an outlook in Section~\ref{outlook}.

\section{Device-independent randomness generation}
\label{diqrng}

%\subsection{Certifying randomness with Bell's Theorem}
DIQRNG protocols make use of the correlations
observed when measuring entangled particles that do not have a
classical analogue, as certified by the violation of a Bell
inequality~\cite{Bell}. The user now needs at least two separated devices for running a
Bell test, see Figure~\ref{fig2}. The devices receive classical
inputs, $x$ and $y$, and produce classical outputs $a$ and
$b$.
After $N_{\rm r}$ rounds of collecting the data $(x,y,a,b)$, the user calculates the relative frequencies of the outcomes given the inputs $P(a,b|x,y)$, which can be estimated without making any assumption about the internal working of the devices.
A Bell inequality is a linear function of these relative frequencies
\begin{equation}
  \label{Bell inequality}
\beta=\sum c_{abxy}\, P(a,b|x,y)
  \leq \beta_L
  \ ,
\end{equation}
characterized by some coefficients $c_{abxy}$.
Here, $\beta_L$ is the so-called
local bound satisfied by classical theories \`a la Einstein-Podolsy-Rosen (EPR)~\cite{epr}. The violation of the Bell
inequality witnesses the presence of non-classical correlations between the two devices.

The idea behind DIQRNG is that if the user observes a Bell
inequality violation, he has the guarantee that the unknown
quantum state in the devices has certain entanglement and purity.
%These two properties, in turn, guarantee the \emph{privacy} and \emph{quality} of the generated outputs.
The purity of the
quantum state certifies that the two devices are not too
correlated with the environment or the external
observer. The entanglement certifies that the local state of
one of the devices is mixed and, thus, a measurement on it
generates random outcomes. Moreover, the Bell
certification of randomness is intrinsically \emph{quantum}, as
classical devices always satisfy a Bell inequality. Finally, it is
\emph{device-independent} as for its computation only the observed
statistics $P(a,b|x,y)$ is needed.

Consider for instance the case in which
the user tests the Clauser-Horne-Shimony-Holt (CHSH) Bell
inequality~\cite{CHSH} and observes its maximal quantum violation\footnote{In what follows some results are illustrated by means of the CHSH inequality, which is the simplest Bell inequality. However, the main ideas and concepts discussed throughout this work apply to any Bell inequality.}.
Take the set of all possible correlations observed after applying local measurements $\Pi_{a|x}$ and $\Pi_{b|y}$ on the state describing the two quantum devices $\ket{\Psi}$,
\begin{equation}
\label{qcorr}
P(a,b|x,y)=\bra{\Psi}\Pi_{a|x}\otimes\Pi_{b|y}\ket{\Psi} .
\end{equation}
Among all these quantum correlations, the only way of getting a maximal violation of the CHSH is when
projective measurements are performed on a maximally entangled state
of two qubits~\cite{Tsirelson,Tsirelson2,PR}
\begin{equation}\label{singlet}
    \ket{\Psi}=\ket{\Phi^+}=\frac{1}{\sqrt 2}(\ket{00}+\ket{11}) .
\end{equation}
As the state is pure, it cannot be correlated to the environment.
The local measurements on half of it produce perfect random
bits~\eqref{rndef}, which are certified by the observed Bell violation. This intuitive argument holds when the maximal quantum violation of the CHSH inequality is obtained. For noisy non-maximal violations, it is also possible to quantify the amount of randomness from the observed violation, see Figure~\ref{fig3}.

%Bell's Theorem states that the outcomes of correlations violating a Bell inequality cannot be determined before the input is applied to the device.
%In a more quantitative fashion, the more the Bell inequality is violated, the less correlated the outcomes can be with information available before the experiment.
%This also limits the degree by which these outcomes can be correlated with the environment after the experiment, since this could be observed beforehand.
%But the crucial property for DIQRNG is that the violation of a Bell inequality is a feature of the statistics of the outcomes and the inputs, independently of how these are generated. %Hence, independently of the functioning of the devices.
%
%This implies that the quality of the randomness is not compromised by the malfunctioning of the devices, nor by the fact that they are maliciously constructed.

While the previous discussion has exploited the properties of quantum correlations~\eqref{qcorr}, it is even possible to design DIQRNG that do not rely on the validity of quantum theory but only on that of the no-signalling principle, that is, the impossibility of faster-than-light communication between devices. In fact, under the sole assumption of no-signalling, the violation of a Bell inequality guarantees the random character of the outputs~\cite{Valentini, BHK05, MAG, bkp, PhDColbeck, Pironio, Hall}.

All these nice features come at a price: the user should meet the conditions needed in~\eqref{qcorr}. %Comparing Figures~\ref{fig1} and~\ref{fig2},
He has to make sure that:

\begin{itemize}
\item (C1): the inputs $(x, y)$ have no correlations with the devices;
\item (C2): there is no communication between the two devices during the generation of the two distant outcomes (see Fig.\ref{fig2}).
\end{itemize}

%\added{CREC QUE ES MES CORRECTE COM VAS ESCRIURE LES DOS CONDICIONS ABANS: (i) there is no communication between the two devices during the generation of the outcomes (see Figure 2), and (ii) the inputs $(x, y)$ have no correlations with the devices. ELS INPUTS NO NOMES HAN DE SER INDEPENDENTS DEL ESTAT, SI NO TAMBE DELS DEVICES. CREC QUE ES MES SIMPLE PARLAR DE DEVICES QUE DE MEASUREMENTS AND STATES. ESTAS D'ACORD?}
%(otherwise they could have been generated by the devices themselves and won't be needed).
Looking at Eq.~\eqref{qcorr}, condition (C1) implies, for instance, that the measured quantum state is independent of $x$ and $y$. Condition (C2)  imposes the tensor product and that the measurements on one device do not depend on the input on the other. To guarantee these conditions the user needs to make physical assumptions on the devices (albeit not on their internal working).
%For example, in some experimental setups, the no-communication requirement can be enforced if the impossibility of faster-than-light communication is assumed.
%These assumptions and DIQRNG protocols are
%discussed next.
%, but notice that already during the previous discussion it was assumed that the devices were completely uncharacterized but quantum. It is possible to relax this assumption and invoke only the validity of the no-signalling principle, which states the impossibility of faster-than-light communication (see Figure~\ref{fig2}). This physical principle is enough for a Bell violation to certify randomness.

%\subsection{Protocols}
\section{Protocols}
\label{protocols}

We now provide a unified description of most of the protocols for DIQRNG proposed so far. %, and compare the assumptions that are required for their security, that is, to guarantee that no observer has knowledge of the generated outputs.
In general, these protocols %require an experimental setup similar to that of a Bell test.
%Hence, they
involve $n \geq 2$ devices, each having an input $x_i$ and an output $a_i$ for $i=1,\ldots, n$.
Condition (C1) imposes that the inputs $(x_1, \ldots, x_n)$ must be selected in a way that is uncorrelated to the devices. A standard, yet not the only, way of satifying this condition is by choosing the inputs using a random {\em seed}, which has to meet some ``independence" requirements depending on the protocol (see below).

%Next, we describe a generic DIQRNG protocol.
\begin{framed}
\noindent
DIQRNG PROTOCOL
\begin{enumerate}

  \item {\bf Data collection.} Repeat $N_{\rm r}$ times steps (a), (b), (c):
  %, and store the string of numbes $r = \{(a_1^{i}, \ldots, a_n^{i}, x_1^{i}, \ldots, x_n^{i})\}_{i=1}^N$. This constitutes the raw data that later will be processed into perfect random bits.
  \begin{enumerate}

  \item A source of $n$-partite entangled states sends a particle to each of the $n$ devices.

  \item Part of the seed is processed to generate a sample from the prior distribution $P (x_1, \ldots, x_n)$ of the inputs applied to each device.
  This distribution can be optimized before the protocol to suit the statistics of the devices.

  \item Measurement $x_i$ is performed on device $i$ generating outcome $a_i$.
  The inputs and outputs of the devices $(a_1, \ldots, a_n, x_1, \ldots, x_n)$ are stored.
  \end{enumerate}

  \item {\bf Non-locality estimation.}
  Calculate the relative frequency of every combination of inputs/outputs $P_{\rm freq} (a_1, \ldots, a_n| x_1, \ldots, x_n)$ using the raw data collected in the $N_{\rm r}$ rounds.
  From this data, estimate the %value of the Bell parameter $\beta$~\eqref{Bell inequality}, and depending on the protocol other parameters as well, and determine the length $N_{\rm k}$ of the final random bit string. The larger the Bell violation, $\beta-\beta_L$, the longer $N_{\rm k}$.
non-locality of the observed correlations and determine the length $N_{\rm k}$ of the final random bit string. The larger the amount of non-locality, the longer $N_{\rm k}$.
If the observed non-locality is insufficient then the protocol is aborted, $N_{\rm k} =0$.

  \item {\bf Classical post-processing.} Generate the final $N_{\rm k}$-bit string using the raw data collected in the $N_{\rm r}$ rounds plus additional part of the seed. This process is often made with a so-called randomness ``extractor"~\cite{randextr,randextr2}.
\end{enumerate}
\end{framed}

A series of parameters that are relevant for the design of DIQRNG protocols are:
\begin{description}

    \item[Efficiency.] Trade-off between the amount of randomness generated and the resources consumed by the DIQRNG protocol. Examples of these resources are the amount of random bits of the seed $N_{\rm s}$ or the number of uses of the devices $N_{\rm r}$.

    \item[Quality of the seed.]
    The random seed may not be perfect. For instance, the seed may not be necessarily uniformly distributed, or display correlations with the devices or the adversary. Another possibility is to assume the existence of a good and free source of public randomness, such as the broadcast by NIST's Randomness Beacon~\cite{NIST}. In this case, the protocol generates private randomness from public randomness. %The advantage of this approach is that it allows using existing public sources of randomness

    \item[Robustness.] Tolerance of the protocol to noise and imperfections. This allows for using realistic noisy apparatuses. %, which are subjected to noise and imperfections.
A protocol is robust if it works ($N_{\rm k} >0$) for violations above some threshold, which does not need to coincide with the local bound.

    \item[Number of devices]used in the protocol. The minimum is $n=2$.

%   \item[Memory effects.] The security does not rely on the ``memoryless" assumption.
%   That is, the $n$ devices can have a malicious internal memory, such that outcomes $(a_1, \ldots, a_n)$ from one round depend on the data $(a_1, \ldots, a_n, x_1, \ldots, x_n)$ from previous rounds.

    \item[Composability.]
    Protocols should be such that, if the adversary learns some information about the $N_{\rm k}$ final random bits, she should be able to deduce essentially no additional information~\cite{ucomp}.

%   \item[Public seed.] Each bit of the  seed is published right after being used, which allows the adversary to choose the measurement on his system depending on the seed.
%   The advantage of this relaxation is that it allows for using existing public sources of randomness~\cite{publicRND} as the seed.

    \item[Physical assumptions.]    Many protocols assume that all the devices and the adversary are constrained by the laws of quantum mechanics. As mentioned, it is possible to relax this requirement and assume %consider protocols where
    only the validity of the no-signalling principle. % is assumed.
%   This case is often called ``no-signaling", although the no-signaling assumption is also necessary when quantum theory is assumed (in order to get the tensor-product structure).
    Even when the security of the protocol does not rely on the validity of quantum mechanics, quantum technology is still needed to generate Bell-violating correlations.
%   A variant of these two possibilities is when the adversary's memory is classical or short term.

\end{description}

A series of works have focused on efficiency, trying to optimize the trade-off between initial and final randomness. The corresponding protocols are known as randomness expansion protocols~\cite{PhDColbeck, Pironio, ColbeckKent, vv, coudron_yuen, miller_shi, Chung, miller_shi2, EATQKD}. Remarkably, it has been proven that an unbounded amount of randomness ($N_{\rm k} \to \infty)$ can be generated from a finite seed~\cite{coudron_yuen, Chung}.

Other works~\cite{coudron_yuen, Chung, crnatphys, rodrigo, BRGHHHSW 15, Bouda, RBHHHW 15, WBGHHHPR 16} have focused on the second point and study how arbitrarily good randomness can be generated in Bell setups using sources of imperfect randomness. These protocols are often known as  {\em randomness amplification} protocols~\cite{crnatphys}. A commonly used model for the imperfect seed $(s_1, s_2, \ldots)$ is a Santha-Vazirani source~\cite{svsource}, which is characterized by a parameter $\epsilon \in (0,1/2)$ such that
    \begin{equation}
      \label{svs}
      \epsilon \leq
      P(s_i|s_1, s_2, \ldots, s_{i-1}, {\rm devices}, {\rm Eve})
      \leq 1- \epsilon\ ,
    \end{equation}
for all $i$.
%  The allowance of correlations between seed, devices and adversary is a new development introduced in~\cite{Chung} and~\cite{WBGHHHPR 16}.
Thus, the larger the value of $\epsilon$, the higher the randomness of the bits. Other models for the seed, more general than the Santha-Vazirani source, have also been considered, such as min-entropy sources~\cite{Chung, Bouda}. In the case of Santha-Vazirani~\eqref{svs}, the performance of randomness amplification protocols is measured by comparing the parameter $\epsilon_{\rm i}$ of the initial source with the final $\epsilon_{\rm f}$ of the generated bits. Full randomness amplification is attained when a source with $\epsilon_{\rm i}\to 0$ is mapped to one with $\epsilon_{\rm f}\to 1/2$~\cite{rodrigo}.   From a fundamental point of view, the existence of protocols attaining full randomness amplification is important; we come back to this point below.
However, from a practical point of view, while allowing for non-perfect seed is a good addition to the security of a protocol, in most applications one can assume that the seed is uncorrelated to the adversary and devices, and the expansion rate is a more practical figure of merit. Randomness expansion and amplification protocols are sides of the general problem, which is the generation of device-independent private randomness under the minimal set of assumptions and with minimal resources (see Table 1).%~\ref{tab:table1}).

\begin{table}[h!]
  \centering
%  \vspace{2mm}
  \label{tab:table1}
  \begin{tabular}{|c||c|c|c|c|c|c|}
    \hline
    &
    &
    &
    & seed
    & seed
    &
    \\     & $n$
    & $N_{\rm k} (N_{\rm s})$
    & $N_{\rm k} /N_{\rm r}$
    & quality
    & privacy
    & QM
    \\ \hline\hline
    Chung...\cite{Chung} &
    large & $\infty$ & $0$ & $\epsilon >0$ & no & yes
    \\ \hline
    Miller...\cite{miller_shi} &
    2 & exp & $>0$ & $\epsilon = 1/2$ & yes & yes
    \\ \hline
%    Brandao... \cite{BRGHHHSW 15} & 4,8 & b & c & weak & secret & no \\ \hline
    Ramanathan...\cite{RBHHHW 15} &
    2 & small & 0 & $\epsilon >0$ & yes & no \\
    \hline
%    Wojew\'odka...\cite{WBGHHHPR 16} &
%    2 & small & 0 & $\epsilon> .49$ & no & yes & no %\\ \hline
    IDEAL &
    2 & $\infty$ & $>0$ & $\epsilon >0$ & no & no \\
    \hline
  \end{tabular}
    \caption{{\bf State of the art in DIQRNG protocols.} Properties of the best known protocols, which are robust and composable. They are also immune to attacks exploiting memory effects on the devices, as in the memory loophole introduced in the context of Bell inequality violations~\cite{memory}. The parameters are, from left to right: number of devices $n$, amount of expansion of the initial seed, efficiency rate, seed quality $\epsilon$ in terms of Eq.~\eqref{svs}, whether the seed can be published without compromising security, and whether the security of the protocol relies on the validity of quantum mechanics (QM). The last row contains the ideal optimal value for each parameter.
}
\end{table}
%As can be seen in Table~\ref{tab:table1}, there are still important challenges. For instance, it would be good to achieve  positive asymptotic rate $N_{\rm k} / N_{\rm r} >0$ from an imperfect seed $\epsilon <1/2$.

Before concluding, it is worth recalling that Bell-certified randomness is also a resource for device-independent quantum key distribution~\cite{BHK05, bkp, EATQKD, Mayers, AGM, acinprl, Masanes, pironio09, hr, MPA, nonfinished, BCK12, RUV12a,BCK12a,VV12}. The goal here, however, is not only to generate randomness, but to establish a secret key between two distant users using a Bell violation observed between their devices. In particular, and in contrast to the case of randomness generation, the devices are held in two separate locations and the channel between them is accessible to the eavesdropper.
%

%\subsection{Implementations}
\section{Implementations}
\label{implementations}

The implementation of the previous DIQRNG protocols requires the observation of a Bell inequality violation.
For that, it is needed to prepare an entangled state of $n\geq 2$  particles, which are distributed to $n$ devices where
they are subjected to local measurements. Assuming the validity of quantum physics, the experimental setup should guarantee that conditions (C1) and (C2) are met so that the observed statistics is correctly described by~\eqref{qcorr}. %The validity of this description is essential for the randomness certification.

An important experimental challenge for the observation of Bell inequality violations is that a high detection efficiency, approximately $\gtrsim 70\%$, is required to close the detection loophole~\cite{pearle}. The loophole says that for low enough detection efficiencies, the statistics of a Bell experiment can always be described by an EPR model, which is deterministic, and thus no randomness certification is possible~\cite{deteff,passaro}. Closing the detection loophole is demanding because it concerns any losses in the setup. But it is a loophole that can be completely, and actually has been closed~\cite{rowe,monroe,weinfurter,zeilinger,kwiat,hanson,giustina,nistexp}. This is because the loophole does not put into question the validity of description~\eqref{qcorr}, but simply demands a high enough detection efficiency.

Contrary to the detection loophole, the locality~\cite{aspect}, collapse-locality~\cite{AK} and free-will~\cite{qrngloop} loopholes do put into question the validity of Eq.\eqref{qcorr}. Because of this, they can never be strictly closed, but their plausibility can be enforced by making physically motivated assumptions on the experimental arrangement. The locality loophole affects condition (C2), that is, whether the measurements in either device %are isolated, i.e., the tensor product in Eq.~\eqref{qcorr} is justified, and
do not depend on what happens on the other device. %, i.e., the operators $\Pi_{a|x}$, resp. $\Pi_{b|y}$, are independent of $y$, resp. $x$.
The standard solution adopted is to invoke Einstein's relativity and arrange the measurements so that they define space-like separated events and no communication can take place between the two devices. This is a very satisfactory solution but also demanding.
%In our view, if the validity of quantum physics is assumed, there are other setups in which, even if it is not possible to completely exclude that some form of communication takes place while the measurements are implemented, the validity of (C2) can be assumed without space-like separated measurements~\cite{Pironio}.
In our view, there are relevant scenarios in which it is also possible to assume the validity of (C2) without space-like separated measurements. For example, some mild level of trust may be put on the provider so that it is safe to assume that the devices do not signal to each other when producing the outputs given the inputs. Or some level of shielding, always essential for any cryptographic use of the generated numbers, may be assumed, which can in turn be used to avoid any unwanted communication among the devices.
%Examples are polarization measurements on two separate photons, or population measurements on two particles in two distant locations.
%Note also that, up to statistical errors, it is always possible to monitor if the no-signalling conditions are satisfied, that is, if the possible unwanted communication has a noticeable effect and the statistics in one device depends on the other. If no signalling is observed, the possible communication between the devices would have to be such that no observable effect is produced, which seems again rather implausible.

Moreover, the space-like arrangement usually adopted to ``close" the locality loophole also assumes that there is a precise knowledge on when the local measurements start and end, that is, when the inputs $x,y$ are defined and the outputs $a,b$ produced. This issue connects with the free-will and collapse-locality loopholes, which also put into question condition (C1). If there is no timing information about when the inputs are generated, this information could for instance exist before the entangled state is produced, i.e., the state in~\eqref{qcorr} could depend on $x$ and $y$. A proposed theoretical solution is that the inputs are generated by human beings, hence the term ``free will". The usual and more practical approach to close the loophole is to use a standard QRNG~\cite{qrngloop}. It is however questionable whether (and if so why) a QRNG is preferable over other processes of classical origin~\cite{Pironioqrng}. Similar considerations apply to the timing of the outputs, as in the collapse-locality loophole: one needs to define when the classical results are actually produced to guarantee that the measurements define space-like separated events.
%{\color{blue} In our view, for many practical applications, it is justified to assume that the initial random seed is not correlated with the quantum devices used.}
%In our view, there are different setups in which it is justified to assume that the choice of settings is not correlated to the devices, not necessarily using a QRNG.

Taking into account all these points and the technological state of the art, there appear two setups in which to implement DIQRNG protocols:  entangled particles in separate locations and entangled photons. In the first case, a Bell test is performed between two distant massive particles, such as nitrogen-vacancy (NV) centres~\cite{hanson}, or ions in two traps~\cite{monroe,weinfurter} that have been entangled through entanglement swapping on two photon-particle entangled pairs~\cite{entsw,simon}. The advantage of this setup is that massive particles can be measured with almost perfect efficiency, thus, closing the detection loophole. In fact, a first proof-of-principle demonstration of DIQRNG, reporting a generation rate of 42 random bits after approximately one month of measurements, was performed using two entangled ions in two traps at 1 meter distance~\cite{Pironio}.
This rate is valid under the assumption that the experimental setup was not operating in a malicious way~\cite{ColbeckKent}.
While challenging, setups with entangled distant particles also allow arranging the measurements so that one can reasonably assume that both the detection and locality loopholes are closed. This was achieved in~\cite{hanson} and subsequently in~\cite{hanson2}.
%The locality loophole can then also be closed, as in the first loophole-free Bell experiment using NV centres~\cite{hanson} or the subsequent experiment~\cite{hanson2}.
However, these experiments do not report any analysis of random-number generation. % and just focused on the observed Bell violation.

The second solution consists of performing polarisation measurements on two entangled photons. Historically, one of the main challenges in these setups was that photon-detection efficiencies were too low to close the detection loophole. However, advances on photo-detectors have allowed closing it~\cite{zeilinger,kwiat}. The first experiment only focused on the Bell violation, but the second reported a random-number generation rate of 0.4 bits/s. More recently, the locality loophole has also been closed in photonic experiments~\cite{giustina,nistexp}%\added{(again invoking some free-will and collapse assumptions)}
, but again none of these experiments were analysed for DIQRNG.

\section{Other methods for randomness generation}
\label{semidiqrng}

Since meeting conditions (C1) and (C2) in an experiment is challenging, alternative proposals for certified quantum randomness generation have been proposed. The idea is to keep part of the device-independent spirit and make only some mild assumptions about the setup, yet without any detailed modelling of the devices. Randomness certification comes from a purely quantum effect with no classical analogue, under the mentioned assumptions. Standard QRNG do not fit into this category, as they require modelling and certify randomness using statistical tests that are also satisfied by classical RNG.

A series of works have explored information protocols under a dimensional constraint~\cite{PB, Pawlowski}, a scenario known as semi-device-independent. The setup is different from a Bell test and consists of a preparing device that prepares a system in different quantum states and a measuring device that performs measurements on it. It is then assumed that the states prepared by the first device and measured by the second belong to a Hilbert space of dimension not larger than $d$. This is the extra assumption that goes beyond the fully DI paradigm. Randomness certification is then obtained via the violation of the so-called dimension witnesseses~\cite{rodrigo2}. One of the practical advantages of this approach is that it %is defined in a prepare-and-measure scenario and
does not require the generation of entanglement. The required detection efficiencies are smaller, but still demanding~\cite{passaro}. A solution to this problem was suggested in~\cite{bowles}, where it was shown that schemes in which one assumes that the preparation and measuring device share no correlations, and that devices do not display memory effects, certify the presence of randomness for any value of the detection efficiency . An proof-of-principle experimental demonstration of this proposal was also performed in~\cite{zbinden}, although a security proof for these schemes without assumptions on memory effects is lacking.

A second proposal considers an asymmetric scenario in which some of the devices are fully trusted. For instance, %in a prepare-and-measure scenario,
one may trust preparing but not the measuring devices. Asymmetric scenarios are often considered in the context of steering~\cite{WJD}. This is a concept defined in the same setup as non-locality, in which two parties perform measurements on two distant quantum particles, but now one of the devices is fully trusted.  The detection of steering provides a quantum certification sufficient to guarantee the presence of randomness~{\cite{branciard}.
Steering has been experimentally demonstrated with the detection and locality loopholes closed~\cite{expsteering1, expsteering2, expsteering3},
%Experimental loophole-free demonstrations of steering have already been reported in
but the detection efficiencies needed for randomness expansion still pose an important challenge~\cite{passaro}. Security proofs are in a preliminary stage also in the case of steering. A general security proof was provided in~\cite{tomamichel}, but it requires a very low level of noise.

\section{Fundamental questions on randomness}
\label{foundations}

The results connecting randomness and non-locality are relevant
not only for applications of quantum technologies,
%protocols for randomness generation,
but also for our %broad
%from a fundamental point of view, as randomness and non-locality play a fundamental role in
understanding of physics.
Protocols attaining full randomness amplification against non-signalling eavesdroppers represent the strongest form of certification using quantum physics of the
existence of random events in nature. It is impossible to certify randomness from scratch. Under only the assumption of no-signalling, the violation of Bell inequalities certifies the presence of randomness, but requires
some initial randomness. Full randomness amplification protocols~\cite{coudron_yuen, Chung, rodrigo, BRGHHHSW 15, Bouda, RBHHHW 15, WBGHHHPR 16} are not able to completely break this circularity, but relax it as much as
possible.

Historically, the whole discussion on EPR models and Bell
inequalities was motivated by the search for a ``complete" alternative
to quantum theory, in the sense that
measurement outcomes could have a deterministic description within
the alternative theory. The violation of Bell inequalities implies that
quantum predictions can not be completed into a deterministic
theory without violating the no-signalling principle. In recent
years, stronger proofs of the ``uncompletability" of quantum theory
have appeared~\cite{BHK05, bkp, crcompl, crcompl1, crcompl2}. These works show that a no-signalling model, possibly non-deterministic, having higher predictive power than quantum theory does not exist.
%These show that quantum predictions not only cannot be completed into a deterministic model, but actually it is impossible to design a no-signalling theory, possibly non-deterministic, that has a higher predictive power than quantum theory for some setups.
%
%
%The results of \cite{bhk,bkp}, for
%instance, obtained in the context of device-independent quantum
%key distribution, show that the measurement outcomes obtained by
%one of the observers when implementing the quantum violation of a
%chained Bell inequality tend to be fully random in the limit of an
%infinite number of measurements in any non-signalling theory, as
%in quantum physics. These results were later extended in~\cite{crcompl1,crcompl2}.
%While these proofs of non-completeness were asymptotic, as they only work number of measurements tend to infinity, another proof of completeness was derived in~\cite{dda} for a simple finite setup including three observers performing two measurements of two outputs\footnote{
Full randomness amplification protocols against no-signalling eavesdroppers can also be seen as proofs for the uncompletability of quantum theory. Along a similar motivation, the study of randomness is also relevant when comparing quantum theory with more general theories respecting the no-signalling principle~\cite{PR,praboxes}. It has been shown that theories leading to general non-signalling correlations do not allow for maximal randomness certification, while quantum theory does~\cite{maxrand}.

Finally, a series of works have shown that the relation between entanglement, non-locality and randomness is
subtler than expected.
%aims at understanding randomness in connection with other quantum resources, such as entanglement and non-locality, and looking for optimal setups for randomness certification. All the works, described in what follows,
For instance, states with arbitrarily
small amounts of entanglement (and non-locality) allow for maximal randomness certification~\cite{AMP}. %One can even wonder whether the it is possible to design situations in which the two bits generated by the two devices define two fully random bits for some choice of measurements, $H_\textrm{glo}=2$.
%Also, it was shown in~\cite{AMP} that two perfect random bits can be certified from a two-qubit  state having arbitrarily small entanglement.
%Later, Bell tests involving $N$ devices and generating $N$ random bits were derived in~\cite{icfo1,icfo2}. Interestingly, this type of maximal randomness certification is impossible in maximally non-local theories limited only by the no-signalling principle~\cite{icfo2}.
Recent progress in this direction also shows that the use of more complex measurements, such as non-projective~\cite{witteketal}
or sequences of measurements~\cite{curchodetal}, provides further advantages for randomness certification.
Despite all these results, a complete understanding of the relation between entanglement, non-locality and randomness is still missing. %This understanding is essential to identify the ultimate limits for randomness certification using quantum resources.}
%Applying general measurements to an entangled state of local dimension $d$, the amount of randomness is upper bounded by $H_\textrm{loc}\leq 2\log_2 d$ and $H_\textrm{glo}\leq 4\log_2 d$, see.
%There, a5Bell test attaining the previous bound on local randomness in the case of qubits was presented, implying the generation of two random bits from a qubit.
%Finally, a more recent work studies a
%slightly different and more complex setup in which sequences of
%(non-destructive) measurements are implemented and shows that in
%this scenario it is possible to certify an unbounded amount of
%randomness starting from a two-qubit
%state~\cite{}.

\section{Outlook}
\label{outlook}

DIQRNG protocols represent a change of paradigm for randomness that solve fundamental and practical drawbacks of standard RNG schemes. On the theory side, the existing security proofs show the validity of the approach. Further theoretical studies are however needed to understand how to relax the requirements for DIQNRG. The ultimate goal would be to design a robust and composably secure protocol attaining an infinite randomness expansion rate using initial sources of arbitrarily weak public randomness with only two devices, and assuming only the validity of the no-signalling principle (see Table~1%\ref{tab:table1}
). While this ambitious goal may be unreachable%~\cite{BCK12a}
, there is still a lot of room for improvement on the conditions needed for DI randomness generation.

On the implementation side, it is expected that new DIQRNG experiments using the setups explained above will be reported in the coming years with a constantly improved generation rate. Looking ahead, integrated photonic circuits and solid-state setups appear as other platforms in which to run the previous protocols. To our knowledge, however, no Bell experiment has been reported on integrated photonic circuits. Bell non-local correlations in solid-state devices have been reported for two ~\cite{martinis,morello} and three systems~\cite{martinis2,dicarlo}. These technologies could be more promising in terms of miniaturisation and, thus, the possible construction of commercial DIQRNG devices. Miniaturisation however comes at a price, as the possible validity of condition (C2) becomes less clear, even if one is ready to accept that measurements don't need to be space-like separated. Theoretical solutions to take into account cross-talk effects have been proposed in~\cite{silman}. In fact, the analysis of DIQRNG implementations opens new theoretical questions, such as: (i) which are the detection efficiencies needed for randomness expansion~\cite{passaro}? (ii) which Bell setups are more robust against noise, or detection inefficiencies~\cite{mattar}? (iii) how to deal with detection inefficiencies~\cite{singapore}?

This is nothing but the natural evolution of this research line, where theory and implementation are joining efforts to design more robust and feasible schemes. RNG with unprecedented standards of quality and security seem within reach using quantum technologies.

\bigskip{\bf Acknowledgments.}
We thank Stefano Pironio, Valerio Scarani, Rotem Arnon Friedman and Andreas Wallraff for discussions.
AA acknowledges support from the ERC CoG QITBOX, the AXA Chair in Quantum Information Science, Spanish MINECO (FOQUS FIS2013-46768-P and SEV-2015-0522), Fundaci\'on Cellex, the Generalitat de Catalunya (SGR 875) and The John Templeton Foundation.
LM is supported by EPSRC.

\section*{Competing financial interests}
The authors declare no competing financial interests.

\newpage

\begin{figure}
\centering
\includegraphics[width=0.65\textwidth]{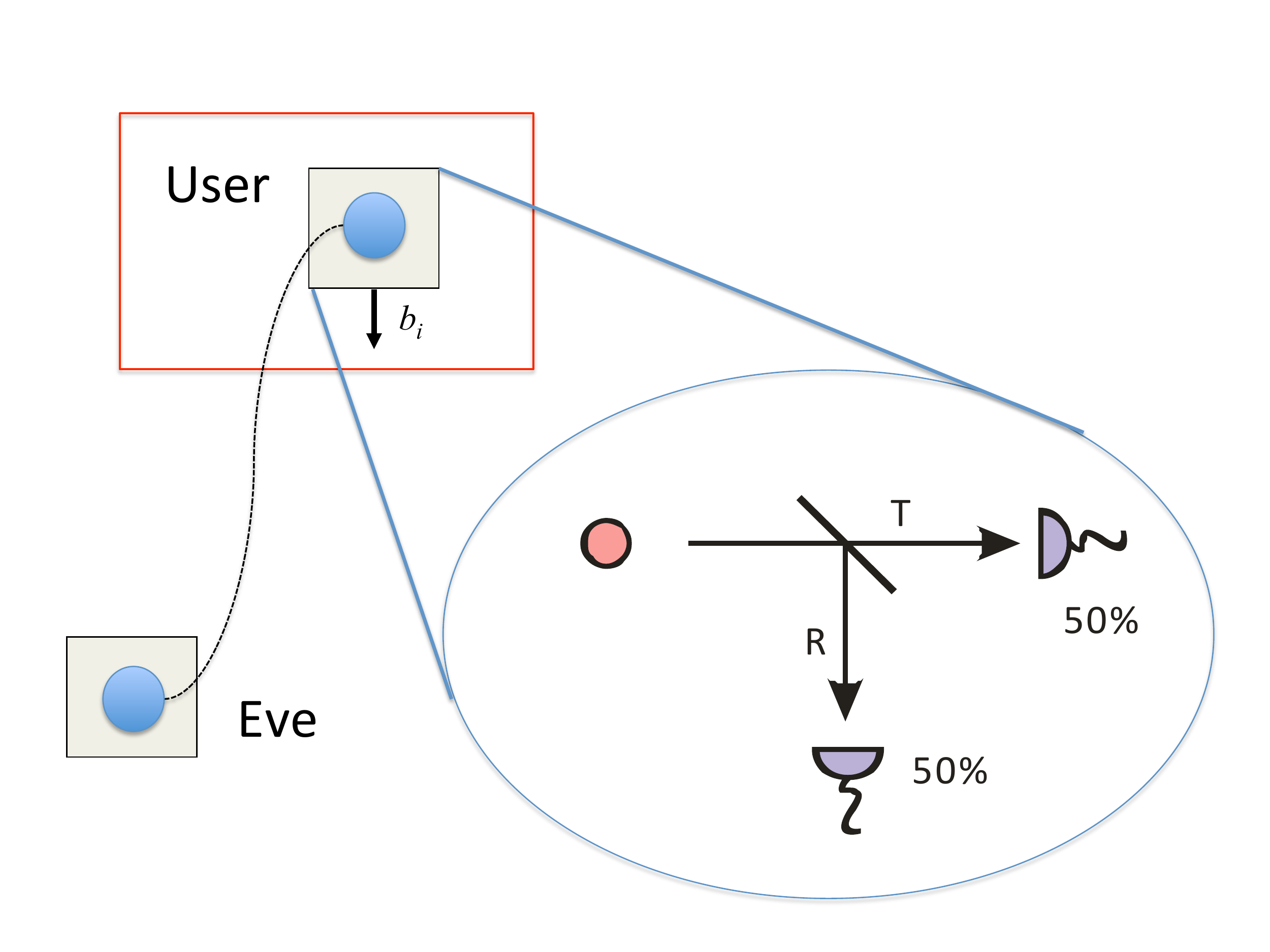}
\caption{\textbf{Schemes for randomness generation.} The user, in
his secure location, has access to a device that generates bits,
$b_i$. The user wants to make sure that the value of these bits cannot
be predicted by any observer outside his lab. The way to model
this is by an external super-observer who has access to all that is
beyond the user's location, represented by another device that may
be correlated to the user's device. It is useful to
interpret the external observer as an adversary or eavesdropper,
Eve, who wants to predict the generated bits (for instance to
break any possible use for cryptographic applications). The
generated bits should be unpredictable to Eve, even after
measuring her device. For standard QRNG, the random character of
the outputs follows from assumptions on the inner working of the
user's device. The figure displays a scheme based on a single
photon (red ball) impinging a beam-splitter with transmission coefficient
equal to $1/2$. Two single-photon detectors placed at the two arms
of the interferometer measure the path taken by the photon.
According to quantum physics, this process is probabilistic and
the probability that a given detector clicks is equal to the
transmission coefficient of the beam-splitter, assumed to be
$1/2$.} \label{fig1}
\end{figure}

\begin{figure}
\centering
\includegraphics[width=0.6\textwidth]{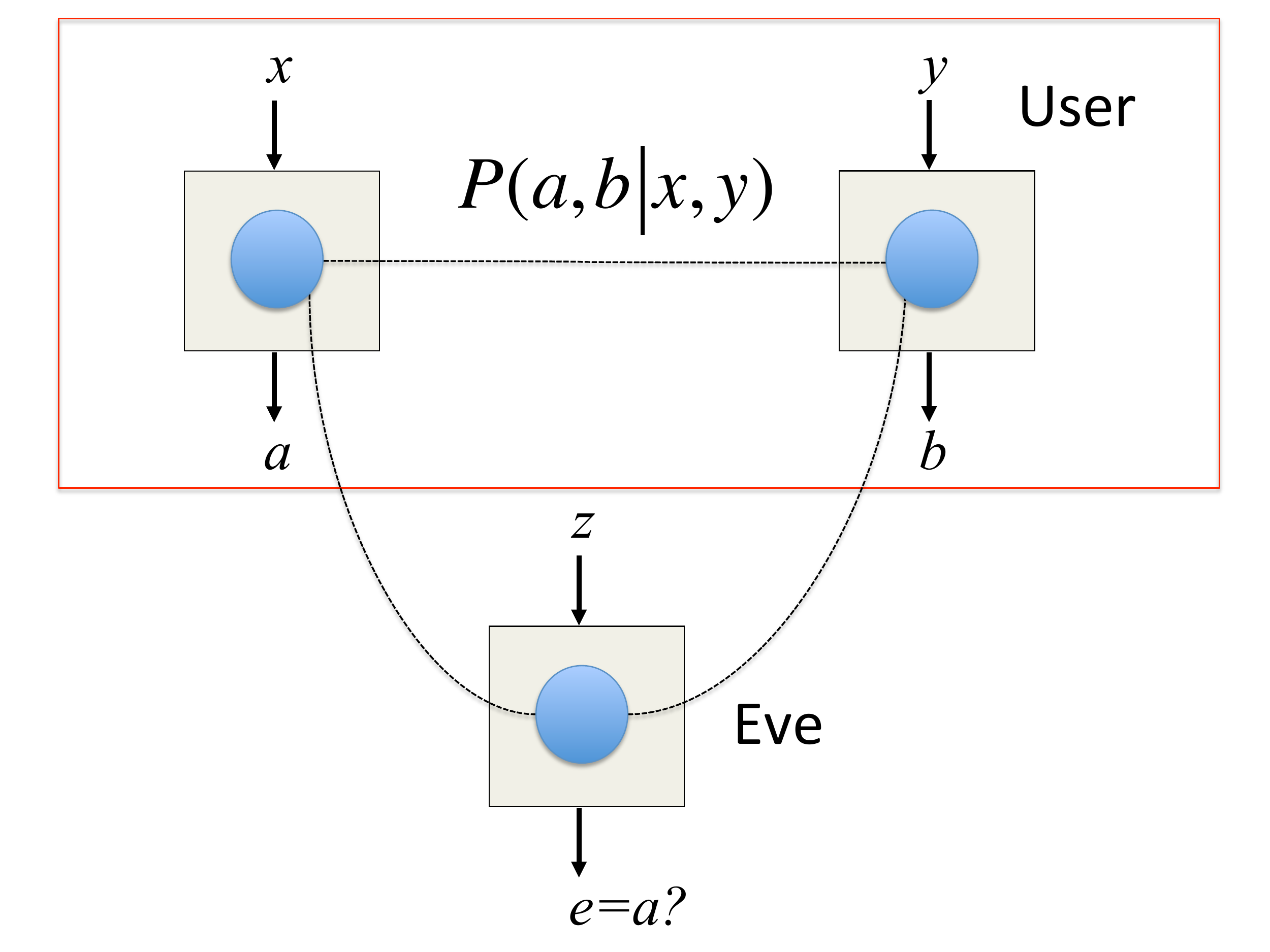}
\caption{\textbf{Structure of DIQRNG protocols.} In a general protocol for DIQRNG the user has access to $n\geq 2$ correlated devices. The figure shows the the simplest case of two devices, which generate classical outputs $a$ and $b$, after applying the inputs $x$ and $y$ (the generalisation to more devices is straightforward).
The inputs $x$ and $y$ can be understood as the labels of the measurement performed on each device
and the outputs as the obtained results.
%The user estimates the relative frequency $P(a,b|x,y)$ of every combination of inputs and outputs.
The external (eavesdropping) observer, Eve, may have
%prepared the devices and may have
 a system correlated with the user's devices.
The randomness of one of the outputs, say $a$, can be quantified by the optimal probability $P_\text{guess}$ that Eve guesses it correctly, $e=a$, after performing a measurement $z$ on her system~\cite{PhDColbeck, Pironio, AMP}.
%This definition can be easily adapted to the two outputs, $a$ and $b$.
In the case of quantum
eavesdroppers, the guessing probability is optimized over all
quantum preparations, including the tripartite state and measurements,
compatible with the correlations observed by the user~\cite{Pironio, NPA, NPA2, Olmo}.
%It is
%possible to bound this quantity~\cite{Pironio,Olmo} using the
%Navascu\'es-Pironio-Ac\'\i n (NPA) hierarchy~\cite{NPA}.
%In addition,
One can
relax the assumption on the validity of quantum mechanics, and consider eavesdroppers who can prepare
any tripartite correlations compatible with the no-signaling
principle, even beyond quantum physics~\cite{Pironio, Olmo}.
%In this supra-quantum case, the guessing probability is in general larger and can be computed using linear
%programming~\cite{Pironio, Olmo}.
} \label{fig2}
\end{figure}

\begin{figure}
\centering
\includegraphics[width=0.8\textwidth]{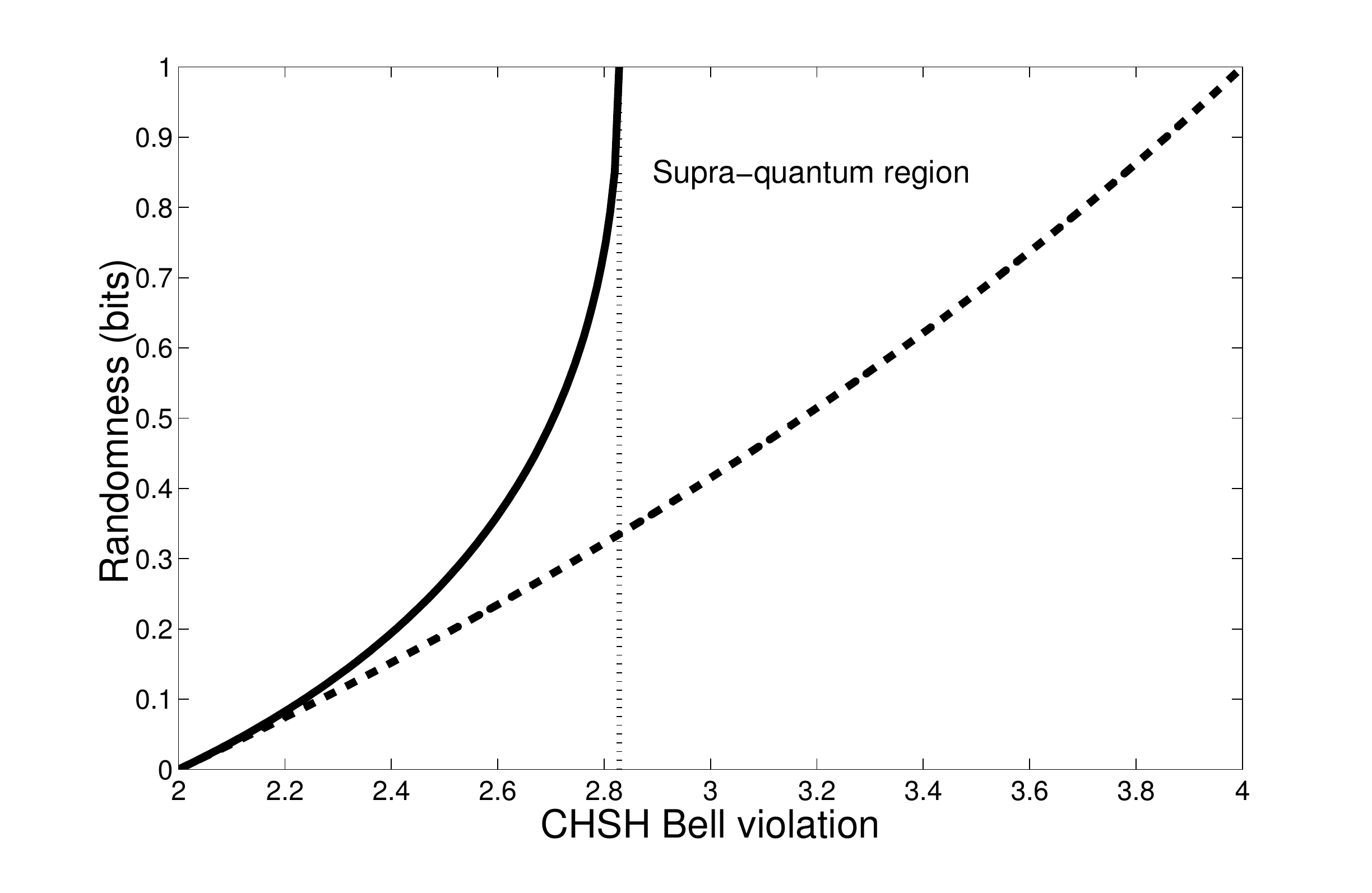}
\caption{\textbf{Randomness for the
CHSH Bell inequality.} Optimal guessing probabilities for one of
the outputs, shown in bits, $-\log_2 P_\text{guess}$, as a
function of the CHSH inequality violation observed by the user.
%For CHSH, measurement outputs take two vales and the local bound is 2.
These curves are computed using the techniques in~\cite{Pironio,AMP}.
The solid line refers to a quantum eavesdropper, while the
dashed line is for a non-signalling eavesdropper. The quantum violation of the CHSH inequality
is upper bounded by $2\sqrt 2$, while it is possible to get larger supra-quantum violations without breaking the no-signalling principle.
At the local bound, CHSH violation equals 2, no randomness can be certified, while some randomness
appears for any non-zero violation. In both cases, a perfect
random bit is certified by the corresponding maximal Bell
violation. %\textbf{AQUESTA PART LA VAIG FER MASSA LLARGA, NO?} \added{ESTIC D'ACORD, PERO ESTA BE EL QUE DIU. SI VOLEM ACURTAR, UNA OPCIO DES TREURE EL BLOC ``In the case of quantum eavesdroppers, the... ...using linear programming [10, 12]."}
} \label{fig3}
\end{figure}

%\newpage

\end{document}